# A Ta-TaS$_2$ monolithic catalyst with robust and metallic interface for superior hydrogen evolution

Qiangmin Yu[1], Zhiyuan Zhang[1], Siyao Qiu[2], Yuting Luo[1], Zhibo Liu[3], Fengning Yang[1], Heming Liu[1], Shiyu Ge[1], Xiaolong Zou[1], Baofu Ding[1], Wencai Ren[3], Hui-Ming Cheng[1, 3, 4], Chenghua Sun[2, 5*], and Bilu Liu[1*]

1. Shenzhen Geim Graphene Center, Tsinghua-Berkeley Shenzhen Institute & Institute of Materials Research, Tsinghua Shenzhen International Graduate School, Tsinghua University, Shenzhen 518055, P. R. China
2. College of Chemical Engineering and Energy Technology, Dongguan University of Technology, Dongguan 523808, P. R. China
3. Shenyang National Laboratory for Materials Sciences, Institute of Metal Research, Chinese Academy of Sciences, Shenyang, Liaoning, 110016, P. R. China
4. Advanced Technology Institute, University of Surrey, Guildford, Surrey GU27XH, UK
5. Department of Chemistry and Biotechnology, and Center for Translational Atomaterials, Swinburne University of Technology, Hawthorn, Victoria 3122, Australia

## Abstract

The use of highly-active and robust catalysts is crucial for producing green hydrogen by water electrolysis as we strive to achieve global carbon neutrality. Noble metals like platinum are currently used in industry for the hydrogen evolution reaction (HER), but suffer from scarcity, high price and unsatisfied performance and stability at large current density, restricting their large-scale implementations. Here we report the synthesis of a new type of monolithic catalyst (MC) consisting of a metal disulfide (e.g., TaS$_2$) catalyst vertically bonded to a conductive substrate of the same metal by strong covalent bonds. These features give the MC a mechanically-robust and electrically near-zero-resistance interface, **leading to an outstanding HER performance including rapid charge transfer and excellent durability, together with a low overpotential of 398 mV to achieve a current density of 2,000 mA cm$^{-2}$ as required by industry. The Ta-TaS$_2$ MC has a negligible performance decay after 200 h operation at large current densities. In light of its unique interface and the various choice of metal elements giving the same structure, such monolithic materials may have broad uses besides catalysis.**

## Introduction

The excessive use of fossil fuel energy has caused serious environmental problems. Hydrogen (H$_2$) is a clean energy carrier with zero-carbon emission and can be produced by water electrolysis driven by renewable energy, which is beneficial for future global carbon neutrality.[1-2] Polymer electrolyte membrane (PEM) electrolyser technology is highly efficient and allows for high hydrogen production rates with current densities up to 2,000 mA cm$^{-2}$, but suffers from problems of poor stability, high cost and low efficiency.[3] Commercial water electrolysis is usually catalyzed by noble metals like platinum (Pt) and iridium (Ir) to produce hydrogen,[4] but these noble metals are scarce and have poor stability especially under large current density.[5-7] Reducing the use of noble metals or developing noble-metal-free catalysts with high activity and durability have been targeted for decades,[8-11] but are far from satisfactory, especially under the large current densities demanded by industry.

Besides large current operation, in practice, stability is another key issue for hydrogen production electrodes, and is usually obtained by anchoring catalysts (such as alloying, clusters, or single-atoms[12-14]) on a conductive substrate using a binder like Nafion. With this approach, the adhesive force is usually weak and the catalysts loaded on the substrate often peel off upon hydrogen bombardment when a large operating current density is used in the hydrogen production, resulting in a short service life of the electrode.[15] Such a structure also inevitably results in a large interface resistance between the catalyst and the substrate, which slows the electron transport and causes serious Joule heating especially at large current densities.[16] As a consequence, the energy conversion efficiency is low, indicating the need to design the catalyst/substrate interface in a conceptually different way. Directly growing the catalyst on a conductive substrate could significantly improve the adhesion between them, improving the robustness of the electrode.[17, 18] Such a technique, however, cannot eliminate the interface resistance, especially when catalyst-substrate interaction is dominated by van der Waals forces or ionic bonds.[19, 20] Therefore, the challenge in producing such an electrode is how to achieve a high-efficiency (ultralow or even zero interface resistance) and long-durability (strong interface binding forces) hydrogen production under large current densities.

Here we develop a monolithic catalyst (MC) to address these challenges. Specifically, a metallic transition metal dichalcogenide (m-TMDC) is vertically grown on a substrate of the same metal using an oriented-solid-phase synthesis (OSPS) method. Due to the nature of the monolith, charges can be directly transferred from the substrate to the catalyst without crossing van der Waals interfaces, providing highly efficient charge injection and an outstanding HER performance. This MC has almost zero interface resistance and therefore offers unimpeded electron transfer. Moreover, the catalyst is bonded to the substrate by covalent bonds, which gives excellent mechanical stability to withstand the large current densities needed for efficient hydrogen production. As an example, a tantalum-tantalum sulfide (Ta-$TaS_2$) MC with a large area has been synthesized by the OSPS method and has shown superior hydrogen evolution activity, achieving 2,000 mA $cm^{-2}$ with a small overpotential of 398 mV and continued working for more than 200 h under large current densities in a 0.5 M $H_2SO_4$ electrolyte without noticeable performance decay.

## Results and discussion

To address the problems of catalyst peel-off and large interface resistance, our strategy is to build the catalyst from the substrate as illustrated in Fig. 1a. Metallic $TaS_2$, the HER catalyst, vertically grows from Ta substrate, with strong Ta-S covalent bonding at their interface. This structure is distinct from normal parallel stacking (see Fig. S1), because there is no van der Waals gap at the interface between Ta and $TaS_2$. Consequently, such a structure fundamentally eliminates catalysts peel-off problem under high-current operations. More importantly, electrons do not have to tunnel over a van der Waals gap between adjacent $TaS_2$ layers to reach active sites. Overall, the design provides an ultra-strong and highly electrically conductive interface for large current density water electrolysis.

For HER, a metal substrate with abundant free electrons could inject electrons into the catalyst effectively for the subsequent reaction, and thus gives the catalyst a high reactivity.[21, 22] The Gibbs adsorption free energy ($G_{H*}$) of Ta-$TaS_2$ MC was calculated. Fig. 1c shows the $G_{H*}$ values of $TaS_2$, Ta-$TaS_2$ MC, and Pt, based on which Ta-$TaS_2$ MC performs similarly to Pt,[23, 24] with $G_{H*}$ ~ 0.10 eV (a $G_{H*}$ value close to zero indicates superior thermodynamic activity[25, 26]), which is much better than $TaS_2$ alone (~ 0.61 eV). The decomposed band structures for hybrid Ta-$TaS_2$ MC were also calculated based on a model (Fig. S1b) and are shown in Fig. 1c. We found that a large number of dispersive electronic states, contributed by Ta and $TaS_2$ jointly, cross the Fermi energy level of the system, confirming that such an interface gives excellent electrical conductivity. To understand the mechanical strength, the energy evolution $E$ with the distance $d$ between Ta and $TaS_2$ has been investigated based on parallel (Ta/$TaS_2$, Fig. S2a) and vertical (Ta-$TaS_2$ MC, Fig. S2b) stacking models. Using respect to equilibrium distance as a reference, energy cost of 0.37 eV and 10.56 eV are resulted to achieve full separation of Ta/$TaS_2$ and Ta-$TaS_2$ MC, respectively, indicating that a mechanical robust interface between Ta and $TaS_2$ has been built in the Ta-$TaS_2$ MC. Accordingly, hybrid Ta-$TaS_2$ MC is expected to exhibit high reactivity, fast kinetics, and strong mechanical stability.

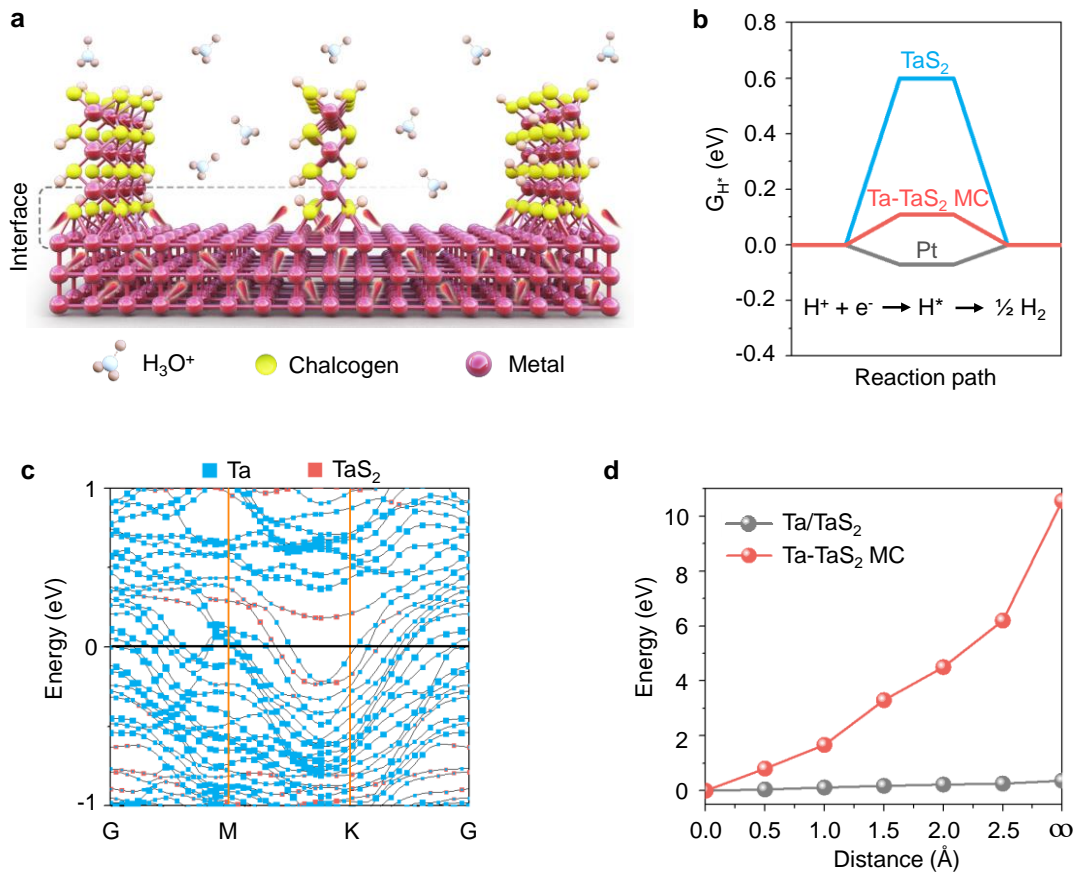

**Figure 1. Structure and properties of the Ta-TaS$_2$ MC material.** (a) Atomic structure of the Ta-TaS$_2$ MC. (b) Hydrogen absorption free energy diagram of TaS$_2$, Ta-TaS$_2$ MC, and Pt catalysts. (c) Its band structures. (d) The separation energies of Ta and TaS$_2$ in parallel Ta/TaS$_2$ and vertical Ta-TaS$_2$ MC materials.

To test the above predictions, we synthesized a Ta-TaS$_2$ MC by an OSPS method and examined its structure. A Ta substrate with periodic holes was pre-oxidized in air (Ta→TaO$_x$), followed by oriented sulfurization along the oxidation path (TaO$_x$→TaS$_2$) and electrochemical treatment to produce a porous MC structure, as illustrated in Fig. 2a and Supplementary Information (Experimental Section). X-ray diffraction (XRD, Fig 2b) indicates the formation of the 3R-phase TaS$_2$ with diffraction peaks of (003) at 14.9°, (101) at 32.2°, and (110) at 55.1° (PDF#89-2756),[27] which has been confirmed by the Raman spectra (Fig. S3).[28] X-ray photoelectron spectroscopy (XPS) measurements show three peaks for the Ta 4f, where the peaks at 23.3 eV and 25.2 eV are assigned to Ta$^{4+}$ in 3R-TaS$_2$ and the one at 27.1 eV is assigned to tantalum oxide. In addition, two S $2p_{3/2}$ (161.9 eV) and S $2p_{1/2}$ (162.9 eV) peaks are assigned to $S^{2-}$ in 3R-TaS$_2$ (Fig. S4).[29] Clearly, a TaS$_2$ 3R phase catalyst had been synthesized using the OSPS method.

A cross-sectional lamella of the Ta-TaS$_2$ MC is shown in Fig. 2c, in which TaS$_2$ vertically growth is seen on a Ta substrate. From high-resolution transmission electron microscopy (HRTEM) images, the TaS$_2$ has an interplanar distance of 0.63 nm, consistent with the (003) plane of the TaS$_2$ 3R-phase (see Figs. 2d and S5).[30] Elemental analysis of the monolithic material by energy dispersive X-ray spectroscopy (EDS) elemental mapping shows that a clear interface was formed between TaS$_2$ and the Ta substrate (Fig. S6). The interface was also examined by scanning TEM-high-angle annular dark field (STEM-HAADF) microscopy (Fig. 2e), with elemental Ta on both two sides while elemental S was present only on one side (Figs. 2f and 2g). The porosity of the MC also can be regulated by laser patterning (Fig. S7), to provide engineerable channels for efficient mass transfer and gas diffusion.[31] From these characterizations, it is clear that a Ta-TaS$_2$ MC has been synthesized. The method can also be used for the synthesis of other MCs, such as niobium-niobium disulfides (Nb-NbS$_2$) and molybdenum-molybdenum disulfides (Mo-MoS$_2$). All three TMDCs have the 3R-phase structure (Figs. S3, S4 and S8).

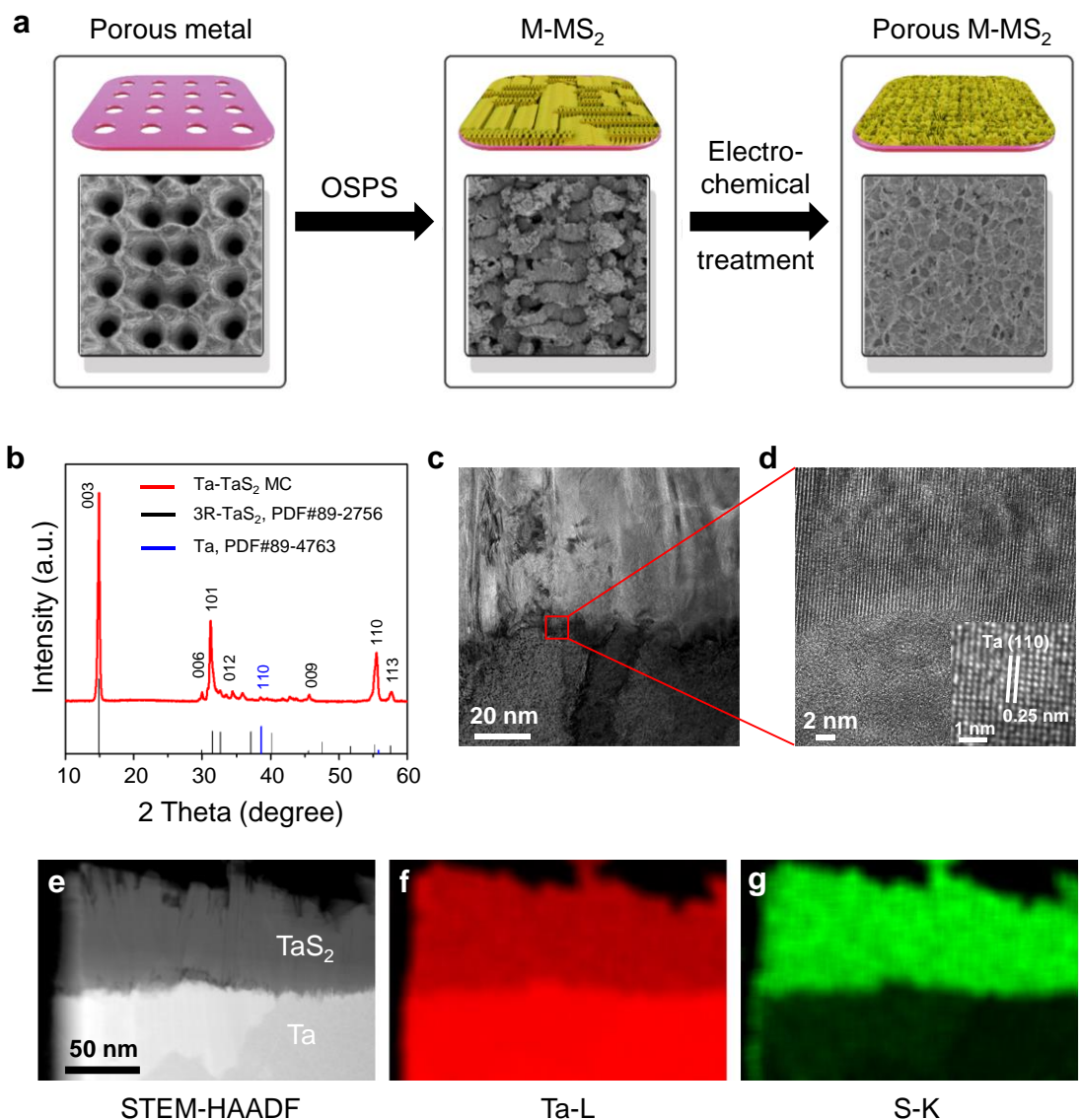

**Figure 2. Synthesis and characterization of the Ta-TaS$_2$ MC.** (a) The OSPS synthesis process of Ta-TaS$_2$ MC and corresponding SEM images. (b) XRD pattern of Ta-TaS$_2$ MC. (c-d) TEM cross-sectional image of Ta-TaS$_2$ MC and a magnified image of the interface. (e) STEM-HAADF image and (f, g) corresponding STEM-EDS elemental maps based on (f) the Ta-L peak and (g) S-K peak.

To investigate the mechanical properties of the interface in Ta-TaS$_2$ MC, a conventional Ta/TaS$_2$ composites (TaS$_2$ catalyst synthesized from Ta oxides loaded on Ta foil, where the catalyst-substrate interface has van der Waals interactions) and a Pt/C/GC composites (Pt/C catalyst pasted on a glassy carbon (GC) substrate using a Nafion binder) were used for comparison. Fig. 3a shows typical force-displacement curves of the Ta-TaS$_2$ MC, and Ta/TaS$_2$ and Pt/C/GC composites. The maximum force at the top of the curve indicates the critical load of the adhesive-bonded joint before a crack starts to propagate.[32] The Ta-TaS$_2$ MC has an adhesive force of 39.9 N/m$^2$, which is more than three times than that of the Ta/TaS$_2$ composites (12.3 N/m$^2$) and the Pt/C/GC composites (13.4 N/m$^2$), indicating a mechanically strong interface in the MC. The electrical conductivity of the Ta-TaS$_2$ MC was examined to investigate its charge transfer ability in HER. As shown in Figs. 3b and S9, an electrical conductivity of ~ 3×10$^6$ S/m was obtained for the Ta-TaS$_2$ material which is comparable to the values for metals (Pt, Ir, Ta) and metallic TaS$_2$. Such a conductivity is 2-5 orders of magnitude higher than those of typical catalysts or substrates including graphite and semiconducting MoS$_2$, and 9 orders of magnitude higher than oxides, indicating excellent charge transfer kinetics across the interface between Ta and TaS$_2$ in the MC, as predicted by theoretical calculations. We also measured the contact angles (CAs) of the Ta-TaS$_2$ MC to analyze its wettability for mass transfer (Fig. S10). The CA is 91.4° for a Ta foil and ~ 0° for the Ta-TaS$_2$ MC, indicating a good wettability of the Ta-TaS$_2$ MC, making it good for mass transfer in an aqueous electrolyte.

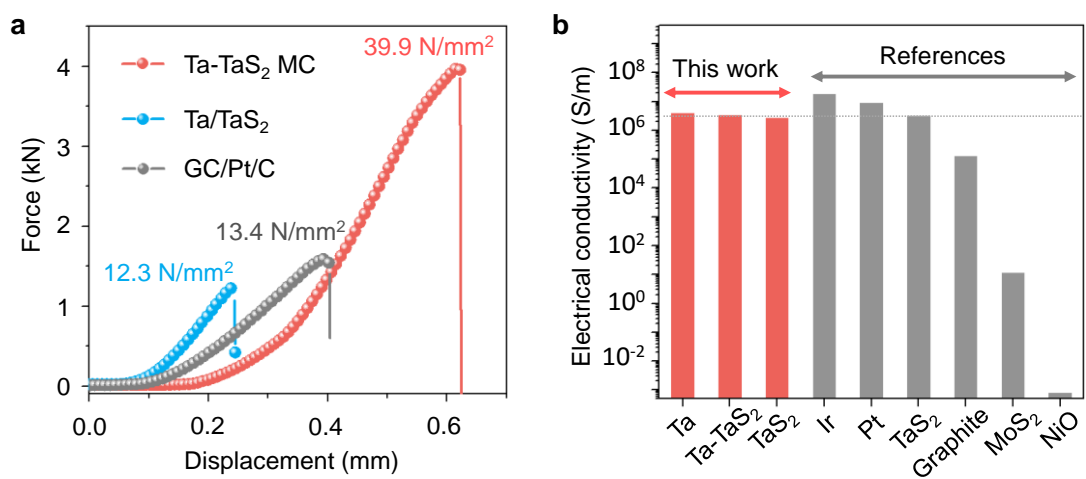

**Figure 3. Mechanical and electrical properties of the Ta-TaS$_2$ MC.** (a) Force-displacement curves for Ta-TaS$_2$ MC, Ta/TaS$_2$, and a commercial Pt/C bound to glassy carbon for comparison. (b) Electrical conductivity of different materials.

Now we turn to examine the HER performance of the MC, in the hope of meeting the needs of large-area synthesis and a high-performance electrocatalyst for large current density ($\geq$1,000 mA cm$^{-2}$) use. The Ta-TaS$_2$ MC has been used as a self-supporting working electrode to evaluate its catalytic performance in a 0.5 M H$_2$SO$_4$ electrolyte (Fig. 4a). It can be clearly seen from the polarization curves that for the current density to reach 2,000 mA cm$^{-2}$ needs an overpotential of only 398 mV, which is much smaller than the value for Ta/TaS$_2$ (920 mV) and a porous Pt foil (740 mV). We used the $\Delta\eta/\Delta\log|j|$ ratios, which has recently been proposed to evaluate the performance of an electrocatalyst over a broad range of current density,[33] to evaluate the performance of catalysts at different current densities (Fig. 4b) and shows that the overpotential increases when the current increases. Both the porous Pt foil and the MC give a low ratio (~ 30 mV dec$^{-1}$) at a small current density, while their responses to current density increases are significantly different. A sharp increase is observed for the porous Pt foil when the current density is larger than 100 mA cm$^{-2}$, and even reaches ~ 90 mV dec$^{-1}$ at 10$^2$-10$^3$ mA cm$^{-2}$. For the Ta-TaS$_2$ MC it remains small, only ~ 58 mV dec$^{-1}$ at 10$^2$-10$^3$ mA cm$^{-2}$, indicating its excellent catalytic performance at large current densities. To verify the effect of the covalently-bonded interface on catalytic performance, Ta/TaS$_2$ has been used as a reference and measured under the same conditions, as shown in Fig. 4c. This shows that the Ta-TaS$_2$ MC always gives a much larger current density than does Ta/TaS$_2$. For example, the current density at an overpotential of 398 mV is 2,000 mA cm$^{-2}$ achieved in the Ta-TaS$_2$ MC, more than three times that with Ta/TaS$_2$ (607 mA cm$^{-2}$). Given that both have TaS$_2$ as the active catalyst, the performance difference is essentially due to the interface, which is not surprising because the HER process at a large current density is overwhelmingly determined by the availability of protons and electrons. As shown above, the MC provides efficient channels for electron transfer between the catalyst and the substrate, which is essential for hydrogen production with a large current density.

The catalytic activity of the MC at a small current density was also measured to evaluate its thermodynamics. Fig. S12 shows the polarization curves of TaS$_2$, Ta-TaS$_2$

MC and porous Pt foil catalysts, according to which the Ta-TaS$_2$ MC has a similar overpotential to Pt at a current density of 10 mA cm$^{-2}$, indicating its high intrinsic activity. In addition, electrochemical impedance spectroscopy curves show that the Ta-TaS$_2$ MC has a charge transfer resistance of 3.2 Ω at an overpotential of 50 mV (Fig S13), notably lower than that of TaS$_2$ (9.0 Ω), which confirms the excellent charge transfer at the covalently-bonded interface. We therefore deduce that the interface in the Ta-TaS$_2$ MC plays a key role in HER at a large current density, which not only demonstrates high catalyst activity, but also provides an unimpeded charge transfer path between the substrate and the catalyst. Similar results have been achieved in other MCs such as Nb-NbS$_2$ and Mo-MoS$_2$ (Figs. S12-14). We also checked the porosity of the MCs (Figs. S15-16) and found that their activities do not have a linear relationship with the numbers of active sites, indicating the key role of the covalently-bonded interface. What more important, the Ta-TaS$_2$ MC is also durable, with no decay even at 1,000 mA cm$^{-2}$ after 200 h operations, as shown in Fig. 4d. Such performance durability has been confirmed by polarization curves and XRD, neither of which show a noticeable change after 20,000 cycles (Figs. S17-19). To give a comprehensive assessment of the MC, the current density values of the Ta-TaS$_2$ MC at η@300 mV have been compared with other state-of-the-art HER catalysts, as shown in Fig. S20 and Table S2. Evidently, the Ta-TaS$_2$ MC, with a current density of 1,120 mA cm$^{-2}$, stands out from them and more importantly, the Ta-TaS$_2$ MC has significant advantages both from large current activity and long-term durability (Fig. 4e and Table S3). These results show that a strong catalyst/substrate interface has been built in the MC, which can support hydrogen production at the large current density required by industry.

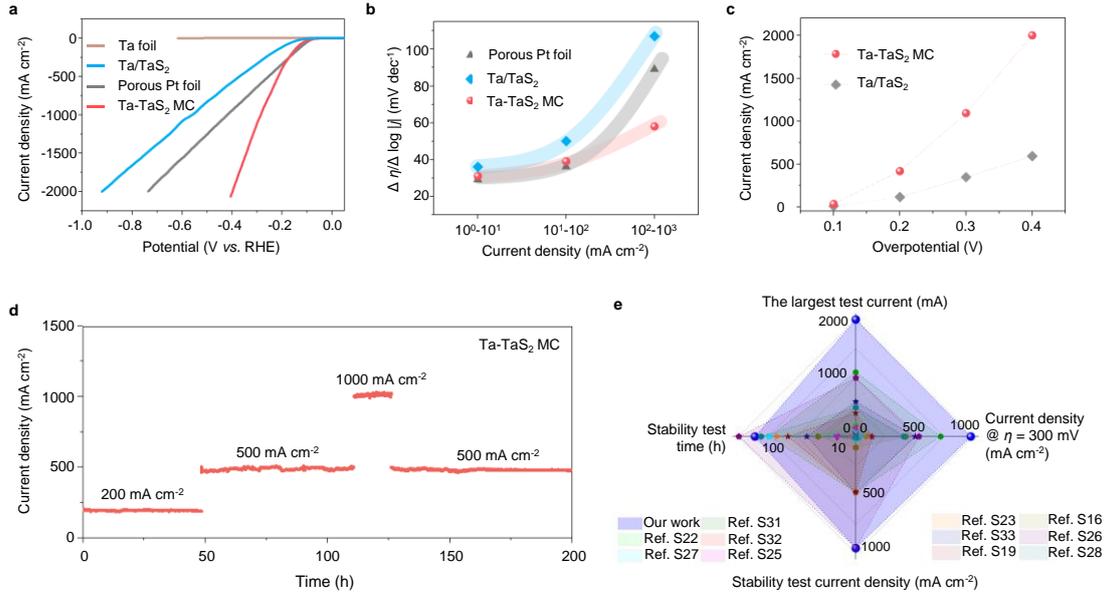

**Figure 4. Large-current-density HER performance of the Ta-TaS₂ MC.** (a) Polarization curves of a Ta foil, Ta-TaS₂ MC, Ta/TaS₂ and a porous Pt foil measured in a 0.5 M $H_2SO_4$ electrolyte with a scan rate of 2 mV s$^{-1}$. (b) Δη/Δlog|$j$| ratios of the Ta-TaS₂ MC, Ta/TaS₂ and porous Pt foil catalysts at different current densities. (c) HER activity of the Ta-TaS₂ MC and Ta/TaS₂. (d) i-t curves of the Ta-TaS₂ MC at various current densities in a 0.5 M $H_2SO_4$ electrolyte. (e) Comprehensive comparisons of the HER performance of Ta-TaS₂ MC with those reported state-of-the-art catalysts in literature.

The production of MC can be scaled-up. As shown in Fig. 5a, a 35 cm² Ta foil, whose size is limited by the diameter of furnace, was used as a precursor to prepare the Ta-TaS₂ MC by the OSPS method. SEM images (Fig. S21) show that the morphology of MC at different regions is similar, indicating good uniformity over a large area. We assembled the Ta-TaS₂ MC as the cathode and commercial iridium oxides ($IrO_2$) as the anode into a home-made electrochemical cell and studied the water electrolysis. Fig. 5b shows that the reaction for the Ta-TaS₂ || $IrO_2$ starts at around 1.50 V and reaches a current density of 1,000 mA cm$^{-2}$ at 1.98 V, which is superior to that of a commercial porous Pt foil || $IrO_2$ couple (2.20 V). $H_2$ and $O_2$ with a volume ratio close to 2:1 were collected in airtight cell (Fig. S22), and the amount of $H_2$ matched well with the calculated results, indicating an almost 100% Faraday efficiency for the HER (Fig. S22).

As for its durability, it is remarkable that this electrolyzer could sustain excellent water-electrolysis with negligible decay for over 24 hours when operating at large current densities of 500 and 1,000 mA cm$^{-2}$. In addition to its catalytic performance, the low cost and abundance of the MC precursors are other advantages for their practical use, as metals like Ta and Nb are 2-3 orders of magnitude cheaper than Pt and their reserves are 1-3 orders of magnitude larger than Pt (Fig. S23), making these MCs extremely promising for industrial hydrogen production by water electrolysis.

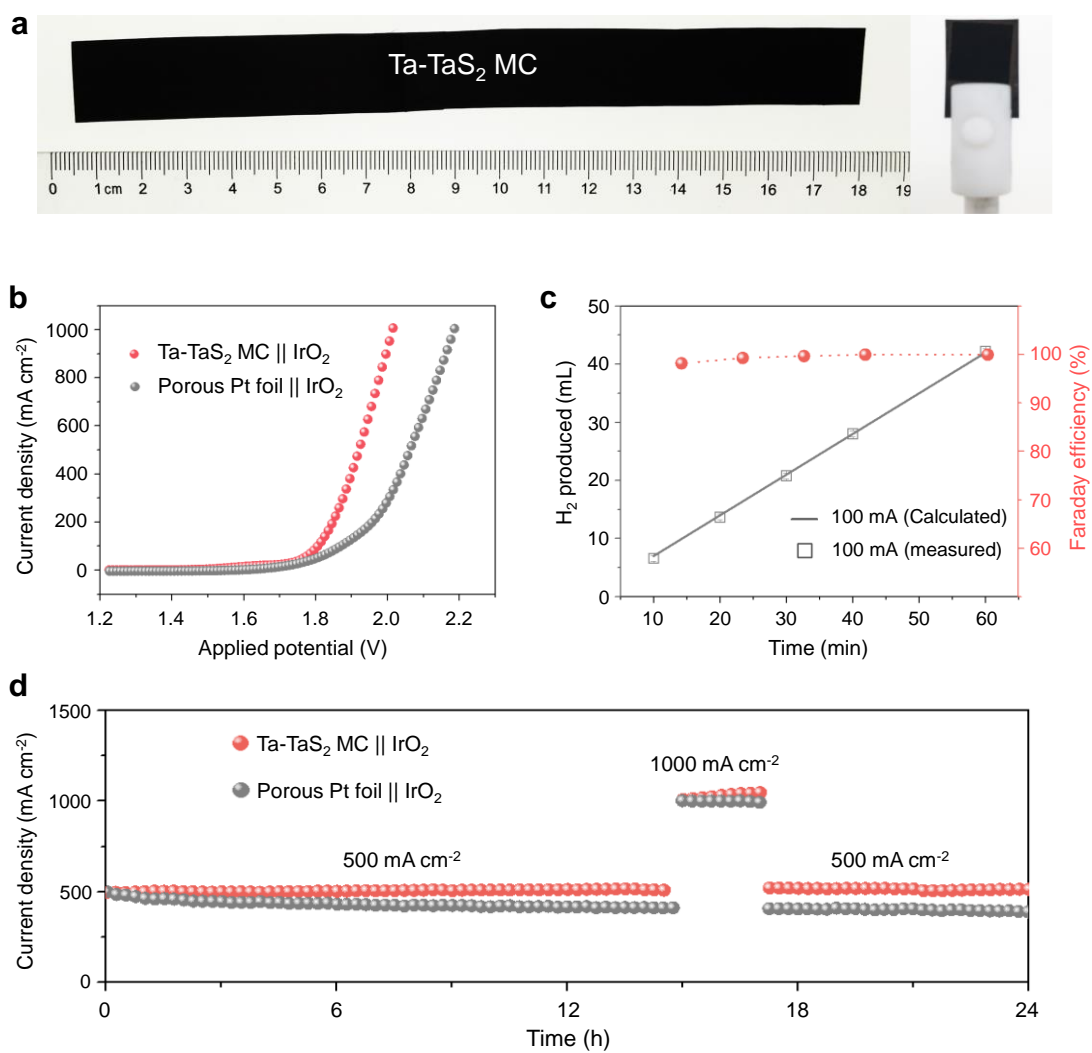

**Figure 5. Scalable synthesis of the Ta-TaS$_2$ MC for PEM water electrolysis.** (a) A photograph of Ta-TaS$_2$, about 175 mm × 20 mm, and corresponding self-supporting electrode. (b) V-I curves of overall water electrolysis with the Ta-TaS$_2$ MC as the cathode and IrO$_2$ as the anode and the current density reaches 1,000 mA cm$^{-2}$. The porous Pt foil ∥ IrO$_2$ were also shown for comparison. (c) Experimental and theoretical

amounts of $H_2$ generated by the Ta-TaS$_2$ electrode at a fixed current density of 100 mA cm$^{-2}$ and corresponding Faraday efficiency. (d) Long-term tests of water electrolysis with Ta-TaS$_2$ or porous Pt foil as the cathodes, and commercial IrO$_2$ as the anode.

## Summary

We have attempted to solve the challenge of large-current-density water electrolysis by the design and the synthesis of MCs. The Ta-TaS$_2$ MC featured a covalently-bonded interface that not only gives it excellent mechanical strength, but also generates outstanding electrical conductivity. As a result, the MC achieves an industrial current density of 2,000 mA cm$^{-2}$, with a small overpotential of 398 mV. It is also durable in a strong acid electrolyte at large current densities for 200 h. For practical use, the MC coupled with commercial IrO$_2$ shows excellent performance in a water electrolyzer, with a HER current density of 1,000 mA cm$^{-2}$ being achieved only by applying a potential of 1.98 V, which is superior to that of commercial Pt and IrO$_2$ couples. The MC can be prepared in large scale and at a low cost, which fills the gap between lab tests and industrial use. Because of the way the material is prepared and its high catalytic performance, the strategy described is this work may be applied to other materials or reactions to solve problems in the energy, chemistry and industrial fields.

**Corresponding authors**
*E-mail: bilu.liu@sz.tsinghua.edu.cn
*E-mail: chenghuasun@swin.edu.au

## Acknowledgements

We acknowledge financial support from the National Natural Science Foundation of China (Nos. 51991340, 51991343, and 51920105002), the Guangdong Innovative and Entrepreneurial Research Team Program (No. 2017ZT07C341), the Guangdong Innovation Research Team for Higher Education (2017KCXTD030), the High-level

## Author contributions

Q.Y., C.S. and B.L. conceived the idea. Q.Y. synthesized the materials, performed characterization, and electrochemical tests. Z.Z., Y.L., F.Y., H.L. and S.G. took part in the electrochemical measurements and discussion. S.Q. and C.S. performed DFT calculations. Z.L. and W.R. performed the TEM characterization and analysis. B.L. supervised the project and directed the research. H.C. advised the project. Q.Y., H.C., C.S. and B.L. interpreted the results. Q.Y., X. Z., H.C., C.S. and B. L. wrote the manuscript with feedback from the other authors.

## Competing interests

Patents related to this research have been filed by Tsinghua-Berkeley Shenzhen Institute, Tsinghua University. The University's policy is to share financial rewards from the exploitation of patents with the inventors.

## Additional information

Supplementary information is available for this paper at…